\documentclass[aps,pra,reprint,superscriptaddress]{revtex4-1}

\usepackage{amssymb, amsmath, graphicx}
\usepackage[usenames, dvipsnames]{color}
\newcommand{\be}{\begin{equation}}
\newcommand{\ee}{\end{equation}}
\newcommand{\ber}{\begin{eqnarray}}
\newcommand{\eer}{\end{eqnarray}}
\newcommand{\de}{\end{equation*}}
\newcommand{\cer}{\begin{eqnarray*}}
\newcommand{\der}{\end{eqnarray*}}

\renewcommand\l{\lambda}

\newcommand\x{\xi}

\newcommand\f{\phi}

\renewcommand\O{\Omega}

\newcommand{\lan}{\langle}
\newcommand{\ran}{\rangle}

\newcommand{\no}{\nonumber}

\newcommand{\BE}{Bose-Einstein~}
\newcommand{\diracslash}[1]{#1\llap{/\kern2pt}}

\begin{document}


\title{On Berezinskii-Kosterlitz-Thouless phase transition and universal breathing mode in two dimensional photon gas}

\author{Vivek M. Vyas}
\affiliation{Department of Physical Sciences, Indian Institute of Science Education \& Research (IISER) - Kolkata, Mohanpur, Nadia - 741252, India}
\affiliation{Theoretical Physics Division, Physical Research Laboratory, Navrangpura, Ahmedabad 380 009, India}

\author{Prasanta K. Panigrahi}
\affiliation{Department of Physical Sciences, Indian Institute of Science Education \& Research (IISER) - Kolkata, Mohanpur, Nadia - 741252, India}

\author{J. Banerji}
\affiliation{Theoretical Physics Division, Physical Research Laboratory, Navrangpura, Ahmedabad 380 009, India}

\date{\today}

\begin{abstract}
A system of two dimensional photon gas has recently been realized experimentally.
It is pointed out that this setup can be used to observe a universal breathing mode of photon gas. It is shown that a modification in the experimental setup would open up a possibility of observing the Berezinskii-Kosterlitz-Thouless (BKT) phase transition in such a system.  It is shown that the universal jump in the superfluid density of light in the output channel can be used as an unambiguous signature for the experimental verification of the BKT transition.
\end{abstract}

\maketitle

\section{Introduction}

It is well known that the Bose-Einstein distribution function has a singularity when the chemical potential reduces to zero.
This singularity can be interpreted as a phase transition, as a result of which at sufficiently low temperature, a system of noninteracting bosons would undergo a Bose-Einstein condensation, which is macroscopic occupation of ground state. Photons are perhaps the best known and certainly most easily accessible bosons around us. It is, therefore, natural to wonder if the photons can undergo \BE condensation. Owing to their vanishing chemical potential, it is easy to show that, in general,  photons would not condense. In 2000, however, a way out was shown in a proposal by Chiao \cite{chiao2000bogoliubov}.
After a decade of unsuccessful attempts, Klaers \emph{et. al.} \cite{klaers2010thermalization,klaers2010} were finally able to realize Chiao's proposal and obtain a \BE condensation of photons.

In the experimental setup of Klaers \emph{et. al.}, the effective dynamics of photons becomes two dimensional. It is well known that such two dimensional systems offer a possibility of infinite order BKT phase transition. However, in the setup of Klaers \emph{et. al.}, there exists an effective harmonic trapping potential for photons that makes the system inhomogeneous and of finite extent. It is known that no true phase transition can exist in such systems even though one can expect to see a broad crossover from one phase to another. A number of works have dealt with the BKT crossover in such finite systems \cite{dalibard,rath,simula,bisset,holzmann}. Such a crossover has also been seen in ultra cold atom experiments \cite{Hadzibabic}.  However, in an interacting Bose system confined in a trap, the distinction between \BE condensate (BEC) transition and BKT transition is not very clear. The problem may partly be attributed to the absence of a clear signature of BKT transition in such systems.

In this paper, we show that a simple modification in the setup of Klaers \emph{et. al.}, would make it possible to observe the BKT phase transition in the system, manifested by the occurrence of a universal jump  in the superfluid density of light. This universal jump is an unambiguous signature of the BKT transition and we show that it can be observed in experiments using interferometry. Furthermore, we point out that the existing setup of Klaers \emph{et. al.} is capable of realizing a universal collective breathing mode of trapped photon gas.

The paper is organized as follows. In the next section, we briefly review the experimental setup and results obtained by Klaers \emph{et. al.}. In the subsequent section, we show how modification of this setup can lead to BKT phase transition. Followed by this is a brief discussion on universal breathing mode. Finally we review obtained results in the last section.

\section{Brief review}

The experimental setup of Klaers \emph{et. al.} consists of an optical micro-cavity filled with a dye solution. The optical cavity is formed by two spherically curved mirrors, and its dimensions are chosen such that the frequency spacing between adjacent longitudinal modes was large and of the order of the spectral width of dye. This leads to modification in dye emission such that the photons trapped inside the cavity predominantly occupy a fixed longitudinal mode. Without loss of generality, it can be assumed that the cavity is along the Z axis; the cavity boundary condition then implies
\be
k_{z}(r) = q \pi/D(r),
\ee
where mirror separation at distance $r$ from the optical (Z) axis is $D(r)=D_{0} - 2 (R - \sqrt{R^{2} + r^{2}})$, $D_{0}$ is the distance ($\simeq 1.46 \: \mu m$) between the two mirrors on the optical axis and $R$ is the radius of curvature ($\simeq 1 \: m$) of mirrors. The longitudinal mode constant $q$, in the above expression, is an integer. Owing to interaction with the dye molecules, photons got locked in the mode $q=7$. So these trapped photons only possess two transverse modal degrees of freedom. In the paraxial approximation (\emph{i.e.,} $k_{z} \gg k_{r} = \sqrt{k^{2}_{x} + k^{2}_{y}}$), their dispersion becomes non-relativistic with boundary conditions manifesting in the form of a harmonic trapping potential:
\begin{align}
E \simeq m_{ph} c^{2} + \frac{(\hbar k_{r})^{2}}{2 m_{ph}} + \frac{m_{ph} \O^{2} r^{2}}{2},
\end{align}
where $m_{ph} = \hbar k_{z}(0)/c$ is the effective photon mass, and $\O = c \sqrt{2/D_{0}R}$ is the effective photon trapping frequency. Interaction of photons with dye molecules introduces a weak nonlinear term in photon dispersion, which can be written in terms of the intensity dependent refractive index $n(r) = n_{0} + n_{2} I(r)$ (here,  $I(r)$ is the optical intensity):
\be \label{disf}
E \simeq m_{ph} c^{2} + \frac{(\hbar k_{r})^{2}}{2 m_{ph}} + \frac{m_{ph} \O^{2} r^{2}}{2} - m_{ph} c^{2}\frac{n_{2}}{n_{0}} I(r).
\ee
Therefore the low energy dynamics of photon gas trapped inside the cavity, is identical to a system of non-relativistic bosons with mass $m_{ph}$, albeit restricted on a plane with harmonic confinement and interacting with each other with a contact potential \cite{dalibard,klaers2010}. It was conclusively shown that the photon gas, which was in contact with the dye solution, was in thermal equilibrium with the solution \cite{klaers2010thermalization}, and the whole system was maintained at room temperature. Since the spacing between adjacent longitudinal modes was much larger than the available thermal energy at room temperature, the average number of photons in the cavity remained conserved. In order to compensate for various kinds of losses, photons were pumped from an external laser source into the cavity. It was found that the system undergoes a phase transition to give rise to a Bose-Einstein condensate, when the total number of photons inside the cavity became larger than the critical number $N_{c} \approx 77,000$.

An unambiguous confirmation that \BE condensation of photons had taken place came from the observation of clear interference patterns, when the photons that were allowed to escape from one of the cavity mirrors were subjected to a Michelson interferometer. This indicated that the photons inside the cavity were coherent. The experimental data obtained in this condensed phase was seen to agree with a theoretical model based on the mean field Gross-Pitaevskii theory, which obeys the  above dispersion relation (\ref{disf}) for certain value of $n_{0}$ and $n_{2}$.

\section{BKT phase transition}\label{proposal}

Consider the case, when the experiment of Klaers \emph{et. al.} is done by using a Fabry-Perot cavity, made of two flat mirrors as originally suggested by Chiao. The quantization condition (1) in such case reads:
\be \no
k_{z} = \frac{q \pi}{D_{0}}.
\ee
Under paraxial approximation, it is easy to see that the dispersion of trapped photons in the cavity, will then become non-relativistic but without a harmonic trapping potential:
\be
E \simeq m_{ph} c^{2} + \frac{(\hbar k_{r})^{2}}{2 m_{ph}} - m_{ph} c^{2}\frac{n_{2}}{n_{0}} I(r).
\ee
In what follows, we shall consider the case when the experiment of Klaers \emph{et. al.} is performed using a Fabry-Perot cavity made out of flat mirrors. Furthermore, we shall assume that one of the mirrors is partially silvered, which allows some photons from the cavity to escape. The same can also be used to pump photons from external source to compensate for various losses. We shall also assume that the transverse dimensions of cavity are large enough so that loss due to the finiteness of the aperture is insignificant.

From the above dispersion,  it is not difficult to see  that the system of photons inside the cavity effectively behaves like a gas of free bosons on a plane with contact interactions. The Hamiltonian describing such a system reads (henceforth we shall work in natural units so that $\hbar=c=k_{B}=1$):
\be
{\mathrm{H}} = \int d^{2}x \: \frac{1}{2m} |\vec{\nabla} {\f}|^{2} + \frac{\pi g'}{2 m} |{\f}|^{4},
\ee
where mass $m=m_{ph}$, coupling constant $g'$ can be expressed in terms of cavity parameters and the boson field $\f$ is the slowly varying component of the electromagnetic field \cite{chiao2000bogoliubov,chiao2004,klaers2010}.

\begin{figure}
\includegraphics[scale=0.4]{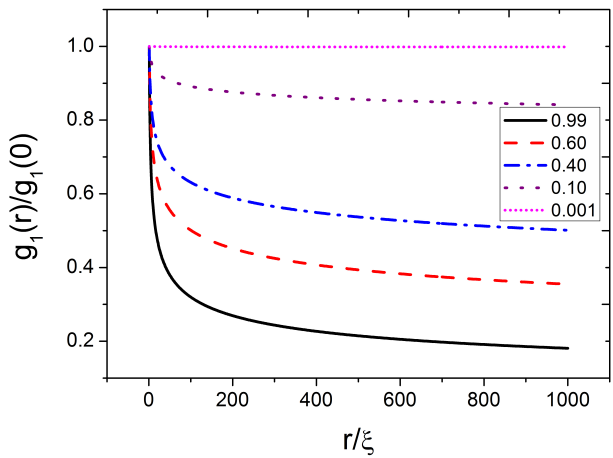}
\caption{\label{g1_lt} (Color online) $\frac{g_{1}(r)}{g_{1}(0)}$ is plotted as a function of $r$ for various values of $T/T_{BKT}$ (with other parameters held fixed), for $T<T_{BKT}$.}
\end{figure}

The problem of interacting bosons in two dimensions is a well-studied one. In this context, a well known result of Bogoliubov states that a system of weakly interacting bosons can undergo a Bose-Einstein condensation only at $T=0$ \cite{bogoliubov1947}. This result seems to forbid the possibility of \BE condensation in such a system. However, it was shown by Berezinskii many years later that, though the system is not a true condensate at finite temperature, it still possesses coherence and exhibit superfluidity \citep{berezinskii}. In particular, it was shown that the first order coherence function for this system reads:
\be \label{ber}
g_{1}(r) = \lan \f(r) \f(0) \ran   \propto \left(\frac{\x}{r} \right)^{\frac{m T}{2 \pi n_{0}}},
\ee
where $n_{0}$ is condensate density (which is also superfluid density) at absolute zero, the constant  $\x = \frac{1}{\sqrt{\pi g' n}}$ is the healing length, and $n$ is number density \cite{berezinskii,dalibard}. Since $g_{1}(r \rightarrow \infty) \rightarrow 0$, it implies that at finite temperature the system is not a true Bose-Einstein condensate (having off-diagonal long range order) as the one at $T=0$, but is a quasi-condensate with phase fluctuations \cite{petrov2003bose}. Although $g_{1}$ vanishes at infinity, as seen in figure (\ref{g1_lt}), the fall as a function of $r$ is slow enough that for all practical purposes one finds $g_{1}$ to be non-zero. Furthermore, it is found that the second order coherence function for this system is close to unity $g_{2}(r) \simeq 1$ \cite{dalibard,petrov2003bose}. This implies that the number density fluctuations in this system at low temperatures are significantly suppressed, and the  system genuinely possesses coherence. The system in such a state is often referred to as having quasi-long range order. Since the photon gas in the cavity possesses coherence, the light coming out of the cavity would also exhibit spatial coherence, with first order coherence function being given by (\ref{ber}).

It is known, however, that apart from phase and density excitations, the system can also exhibit topological excitations like vortices. In the absence of density fluctuation, the dynamics of the system in this (quasi) condensed phase can solely be described in terms of a phase field $\theta$, where $\phi = n_{0} e^{i \theta}$. Since the $\theta$ field satisfies Laplace equation ${\nabla}^{2} \theta(x) = 0$, it admits a singular solution with a branch cut $\theta(r,\varphi) = q \varphi$ (where $q$ is an integer), such that there is non-zero circulation:
\be
\frac{1}{2 \pi} \oint d\vec{l} \cdot \vec{\nabla} \theta = q.
\ee
These are quantized vortex solutions with topological charge $q$.  Energy stored in a single vortex field is given by:
\be
E = \frac{n_{0} \pi}{m} \int_{\x}^{R} \frac{dr}{r} = \frac{n_{0} \pi}{m} \text{ln} \left(\frac{R}{\x} \right),
\ee
where $R$ is the radial extent of the system. As is evident, energy for a single vortex field goes to infinity in the thermodynamic limit, implying that an isolated single vortex is unstable at absolute zero. Entropy to create such a vortex is given by
$S=2\text{ln}(R/\x)$, and hence Helmholtz free energy is given by:
\begin{align}\nonumber
F &=E-TS\\ \no &= \text{ln} \left(\frac{R}{\x} \right) \left( \frac{n_{0} \pi}{m} - 2 T \right).
\end{align}
This implies that as long as $T< {n_{0} \pi}/{2m}$, vortex formation is not favored thermodynamically, however if $T> {n_{0} \pi}/{2m}$ then vortex solution minimizes free energy and hence vortex formation is favored.
In the presence of vortex field $\theta_{v}$ due to a vortex-anti vortex pair, which is the singular part of $\theta$ field, superfluid density $n_{s}$ no longer equals $n_{0}$, and gets diminished \cite{chaikin}:
\be
n_{s} = n_{0} - \frac{n^{2}_{0}}{T} \int d^{2}x \: \lan \vec{\nabla} \theta_{v}(\vec{r}) \cdot \vec{\nabla} \theta_{v}(0)\ran
\ee
where $\lan \cdots \ran$ represent thermal average. This renormalization of density favors creation of more vortex pairs since $T> {n_{0} \pi}/{2m}$ would still hold. As a result one sees that the superfluid state becomes unstable due to vortex proliferation and the system ultimately ends up losing superfluidity. This is the celebrated Berezinskii-Kosterlitz-Thouless (BKT) phase transition \cite{berezinskii,kosterlitz}, which separates superfluid (quasi-coherent) and normal (incoherent) phases of the system, with critical point being:
\begin{equation} \nonumber
T_{BKT}=\frac{n_{0} \pi}{2m}.
\end{equation}
It was shown by Nelson and Kosterlitz \cite{nelson1977}, using the renormalisation group approach, that the superfluid density shows a universal jump around the critical point:
\be \no
n_{s}(T) =\begin{cases} \frac{2 m }{\pi} T_{BKT} & \text{for}\: T \rightarrow T^{-}_{BKT},\\
0&  \text{for}\: T \rightarrow T^{+}_{BKT}.\end{cases}
\ee

\begin{figure}
\includegraphics[scale=0.4]{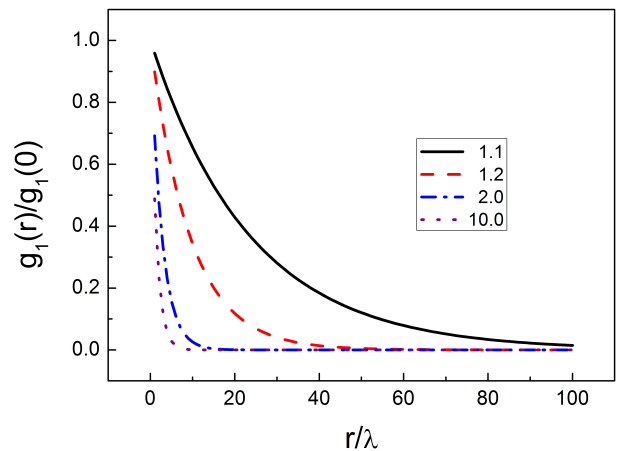}
\caption{\label{g1_ht} (Color online) $\frac{g_{1}(r)}{g_{1}(0)}$ is plotted as a function of $r$ for various values of $T/T_{BKT}$ when $T>T_{BKT}$ (with other parameters held fixed). Note the sharp decline as $r$ increases.}
\end{figure}

The above relation shows that, whatever may be the value of $n_{s}$ at $T=0$, at $T=T^{-}_{BKT}$ superfluid density $n_{s}$ must be $(2 m/\pi) T_{BKT}$. This jump in the superfluid density has been observed in several condensed matter experiments \cite{bishop1978,mondal}. Above BKT critical point, due to the presence of vortices, the quasi-long range order is lost and the first order coherence function shows an exponential decay:
\be \no
g_{1}(r) \propto e^{- r/l},
\ee
where correlation length $l = \l \; \text{exp}\left( \sqrt{\frac{a T_{BKT}}{T - T_{BKT}}}\right)$ (here $a$ is a model dependent constant and $\lambda= \frac{1}{\sqrt{2 \pi m T}}$ is the thermal de-Broglie wavelength), diverges at the critical point \cite{dalibard,chaikin}. This is indicative of loss of coherence
and so the system is no longer in a quasi-coherent state.

In this discussion, we have assumed that density fluctuations are negligible even at BKT transition temperature, and do not affect transition mechanism. Numerical simulations performed by Prokof'ev \emph{et. al.} suggest that the density $n(r)$ at the critical point obeys the following relation \cite{prokofev2001}:
\be
\frac{\sqrt{2 n^{2} - \lan n^{2}(r) \ran}}{n} = \frac{7.16}{\text{ln} (\frac{c}{\pi g'})},
\ee
where $c=380 \pm 3$ and $n = \lan n(r) \ran$. One finds for realistic values of $g'$, that $\lan n^{2} \ran$ is of the order of $\lan n \ran^{2}$, implying that density fluctuations are suppressed even at BKT transition point.

Above we saw that critical temperature depends on condensate density at absolute zero $n_{0}$. However it is desirable that one knows accurately critical temperature in terms of the number density, which can be controlled in experiments. After significant analytical and numerical efforts, relation between critical temperature and number density was found to be \cite{prokofev2001,fisher1988}:
\be
T_{BKT} = \frac{n}{2 \pi m \: \text{ln}\left( \frac{c}{\pi g'} \right)},
\ee
when $g' \ll 1$. It is thus evident that $T_{BKT}$ can be controlled by varying cavity parameters and more easily, by changing the number density of photons in the cavity. Considering values of cavity parameter to be $D_{0} \approx 1 \, \mu m$,  coupling constant $g'\approx 10^{-4}$ and $n \approx 10^{11} \, m^{-2}$, which are realized in the experiments by Klaers \emph{et. al.}, one finds that $T_{BKT} \approx 0.1 \, K$, which is achievable in laboratory. On the other hand, if one has $n \approx 10^{14} \, m^{-2}$ then critical temperature happens to be near room temperature $T_{BKT} \approx 300 \, K$.

It is not difficult to see that, unlike condensed matter systems, measurement of superfluid density is challenging in these systems. There have been a few proposals to directly measure superfluid density exploiting phenomenological properties exhibited by a superfluid \cite{bolda,leb}. In 1966, Josephson showed a beautiful connection between superfluid density and first order coherence function (or two- point correlation function) in a given system \cite{Jos}, which for our case reads:
\begin{equation}
n_{s} =  - \lim_{k \to 0} \frac{n_{0} m^{2}}{ k^{2} G_{1}(k)},
\end{equation}
where $G_{1}(k)$ is Fourier transform of $g_{1}(r)$. This relation is particularly interesting as it provides an indirect route to determine superfluid density, since first order coherence function $g_{1}(r)$ can be measured in experiments via interferometry \cite{mandel}, from which one can determine $n_{0}$ using equation (\ref{ber}).

\section{Universal breathing mode}

Above we saw that collective behaviour of photons can exhibit an infinite order phase transition of unique kind. This may not happen in the case of experiments by Klaers \emph{et. al.}, where spatially inhomogeneous and finite system is realized. It is known that such systems exhibit smooth crossovers rather than phase transitions.
However, as we point out below, such a system would exhibit a unique collective mode.

The problem of  non-relativistic bosons, restricted in two dimensions, with contact interactions and a harmonic trap has been studied by Pitaevskii more than a decade ago \cite{Pitaevskii1997}. He showed, in several ways, that there is a hidden SU(1,1) symmetry in the theory, which is responsible for the existence of a universal breathing mode in the system. It was shown clearly that, if the initial state of the system is out-of-equilibrium, then the system exhibits an undamped breathing mode with frequency twice that of harmonic trap, independent of other parameters in the theory. In other words, if  $X$ is the mean square displacement of particles, then it obeys the relation
\be
X(t) = X_{0} + A\: \text{cos}(2 \O t),
\ee
where $\O$ is the frequency of harmonic trap, $X_{0}$ and $A$ are constants independent of other parameters in the system. Subsequently, this universal breathing mode was observed in many experiments with ultra cold atoms  \cite{Bloch2008}. Since the photons in the experiments of Klaers \emph{et. al.} are well described by equation (\ref{disf}), Pitaevskii's argument holds and the system under suitable circumstances should show the presence of these collective breathing modes. This mode can be excited by sending a sharp intense light pulse in the cavity, which can drive the system away from equilibrium. An experimental signature of these modes would be a change in spatial intensity distribution (as a function of time) of the light that is emanating out of the cavity, with a periodicity twice that of the harmonic trap.

\section{Conclusion}

In conclusion, it is pointed out that the experiment by Klaers \emph{et. al.}, when done using flat mirrors can show the occurrence of BKT transition in the system. It is necessary, however, that the transverse dimensions of the system are much larger than the relevant length scale, which in this case is the healing length, so that the effects due to the finiteness of the system are unimportant. As shown above, an unambiguous observation of the BKT transition is the
universal jump in the superfluid density of output light as the critical point is traversed. The possibility of BKT transition in photon gas has been considered in \cite{chiao2004} but the critical temperature was found to be too high to attain in laboratory. In the current proposal, critical temperature is found to be well within the reach of experiments, and since we rely on a scheme which has already yielded BEC, we believe there should not be any other hurdle in observing the BKT transition. It remains to see if such a phase transition is realized in future experiments.


\begin{thebibliography}{26}%
\makeatletter
\providecommand \@ifxundefined [1]{%
 \@ifx{#1\undefined}
}%
\providecommand \@ifnum [1]{%
 \ifnum #1\expandafter \@firstoftwo
 \else \expandafter \@secondoftwo
 \fi
}%
\providecommand \@ifx [1]{%
 \ifx #1\expandafter \@firstoftwo
 \else \expandafter \@secondoftwo
 \fi
}%
\providecommand \natexlab [1]{#1}%
\providecommand \enquote  [1]{``#1''}%
\providecommand \bibnamefont  [1]{#1}%
\providecommand \bibfnamefont [1]{#1}%
\providecommand \citenamefont [1]{#1}%
\providecommand \href@noop [0]{\@secondoftwo}%
\providecommand \href [0]{\begingroup \@sanitize@url \@href}%
\providecommand \@href[1]{\@@startlink{#1}\@@href}%
\providecommand \@@href[1]{\endgroup#1\@@endlink}%
\providecommand \@sanitize@url [0]{\catcode `\\12\catcode `\$12\catcode
  `\&12\catcode `\#12\catcode `\^12\catcode `\_12\catcode `\%12\relax}%
\providecommand \@@startlink[1]{}%
\providecommand \@@endlink[0]{}%
\providecommand \url  [0]{\begingroup\@sanitize@url \@url }%
\providecommand \@url [1]{\endgroup\@href {#1}{\urlprefix }}%
\providecommand \urlprefix  [0]{URL }%
\providecommand \Eprint [0]{\href }%
\providecommand \doibase [0]{http://dx.doi.org/}%
\providecommand \selectlanguage [0]{\@gobble}%
\providecommand \bibinfo  [0]{\@secondoftwo}%
\providecommand \bibfield  [0]{\@secondoftwo}%
\providecommand \translation [1]{[#1]}%
\providecommand \BibitemOpen [0]{}%
\providecommand \bibitemStop [0]{}%
\providecommand \bibitemNoStop [0]{.\EOS\space}%
\providecommand \EOS [0]{\spacefactor3000\relax}%
\providecommand \BibitemShut  [1]{\csname bibitem#1\endcsname}%
\let\auto@bib@innerbib\@empty
\bibitem [{\citenamefont {Chiao}(2000)}]{chiao2000bogoliubov}%
  \BibitemOpen
  \bibfield  {author} {\bibinfo {author} {\bibfnamefont {R.}~\bibnamefont
  {Chiao}},\ }\href@noop {} {\bibfield  {journal} {\bibinfo  {journal} {Opt.
  Comm.}\ }\textbf {\bibinfo {volume} {179}},\ \bibinfo {pages} {157} (\bibinfo
  {year} {2000})}\BibitemShut {NoStop}%
\bibitem [{\citenamefont {Klaers}\ \emph
  {et~al.}(2010{\natexlab{a}})\citenamefont {Klaers}, \citenamefont
  {Vewinger},\ and\ \citenamefont {Weitz}}]{klaers2010thermalization}%
  \BibitemOpen
  \bibfield  {author} {\bibinfo {author} {\bibfnamefont {J.}~\bibnamefont
  {Klaers}}, \bibinfo {author} {\bibfnamefont {F.}~\bibnamefont {Vewinger}}, \
  and\ \bibinfo {author} {\bibfnamefont {M.}~\bibnamefont {Weitz}},\
  }\href@noop {} {\bibfield  {journal} {\bibinfo  {journal} {Nature Physics}\
  }\textbf {\bibinfo {volume} {6}},\ \bibinfo {pages} {512} (\bibinfo {year}
  {2010}{\natexlab{a}})}\BibitemShut {NoStop}%
\bibitem [{\citenamefont {Klaers}\ \emph
  {et~al.}(2010{\natexlab{b}})\citenamefont {Klaers}, \citenamefont {Schmitt},
  \citenamefont {Vewinger},\ and\ \citenamefont {Weitz}}]{klaers2010}%
  \BibitemOpen
  \bibfield  {author} {\bibinfo {author} {\bibfnamefont {J.}~\bibnamefont
  {Klaers}}, \bibinfo {author} {\bibfnamefont {J.}~\bibnamefont {Schmitt}},
  \bibinfo {author} {\bibfnamefont {F.}~\bibnamefont {Vewinger}}, \ and\
  \bibinfo {author} {\bibfnamefont {M.}~\bibnamefont {Weitz}},\ }\href@noop {}
  {\bibfield  {journal} {\bibinfo  {journal} {Nature}\ }\textbf {\bibinfo
  {volume} {468}},\ \bibinfo {pages} {545} (\bibinfo {year}
  {2010}{\natexlab{b}})}\BibitemShut {NoStop}%
\bibitem [{\citenamefont {{Hadzibabic}}\ and\ \citenamefont
  {{Dalibard}}(2009)}]{dalibard}%
  \BibitemOpen
  \bibfield  {author} {\bibinfo {author} {\bibfnamefont {Z.}~\bibnamefont
  {{Hadzibabic}}}\ and\ \bibinfo {author} {\bibfnamefont {J.}~\bibnamefont
  {{Dalibard}}},\ }\href@noop {} {\bibfield  {journal} {\bibinfo  {journal}
  {arXiv}\ } (\bibinfo {year} {2009})},\ \Eprint
  {http://arxiv.org/abs/0912.1490} {0912.1490 [cond-mat.quant-gas]}
  \BibitemShut {NoStop}%
\bibitem [{\citenamefont {Rath}\ \emph {et~al.}(2010)\citenamefont {Rath},
  \citenamefont {Yefsah}, \citenamefont {G\"unter}, \citenamefont {Cheneau},
  \citenamefont {Desbuquois}, \citenamefont {Holzmann}, \citenamefont
  {Krauth},\ and\ \citenamefont {Dalibard}}]{rath}%
  \BibitemOpen
  \bibfield  {author} {\bibinfo {author} {\bibfnamefont {S.~P.}\ \bibnamefont
  {Rath}}, \bibinfo {author} {\bibfnamefont {T.}~\bibnamefont {Yefsah}},
  \bibinfo {author} {\bibfnamefont {K.~J.}\ \bibnamefont {G\"unter}}, \bibinfo
  {author} {\bibfnamefont {M.}~\bibnamefont {Cheneau}}, \bibinfo {author}
  {\bibfnamefont {R.}~\bibnamefont {Desbuquois}}, \bibinfo {author}
  {\bibfnamefont {M.}~\bibnamefont {Holzmann}}, \bibinfo {author}
  {\bibfnamefont {W.}~\bibnamefont {Krauth}}, \ and\ \bibinfo {author}
  {\bibfnamefont {J.}~\bibnamefont {Dalibard}},\ }\href@noop {} {\bibfield
  {journal} {\bibinfo  {journal} {Phys. Rev. A}\ }\textbf {\bibinfo {volume}
  {82}},\ \bibinfo {pages} {013609} (\bibinfo {year} {2010})}\BibitemShut
  {NoStop}%
\bibitem [{\citenamefont {Simula}\ \emph {et~al.}(2008)\citenamefont {Simula},
  \citenamefont {Davis},\ and\ \citenamefont {Blakie}}]{simula}%
  \BibitemOpen
  \bibfield  {author} {\bibinfo {author} {\bibfnamefont {T.~P.}\ \bibnamefont
  {Simula}}, \bibinfo {author} {\bibfnamefont {M.~J.}\ \bibnamefont {Davis}}, \
  and\ \bibinfo {author} {\bibfnamefont {P.~B.}\ \bibnamefont {Blakie}},\
  }\href@noop {} {\bibfield  {journal} {\bibinfo  {journal} {Phys. Rev. A}\
  }\textbf {\bibinfo {volume} {77}},\ \bibinfo {pages} {023618} (\bibinfo
  {year} {2008})}\BibitemShut {NoStop}%
\bibitem [{\citenamefont {Bisset}\ \emph {et~al.}(2009)\citenamefont {Bisset},
  \citenamefont {Davis}, \citenamefont {Simula},\ and\ \citenamefont
  {Blakie}}]{bisset}%
  \BibitemOpen
  \bibfield  {author} {\bibinfo {author} {\bibfnamefont {R.~N.}\ \bibnamefont
  {Bisset}}, \bibinfo {author} {\bibfnamefont {M.~J.}\ \bibnamefont {Davis}},
  \bibinfo {author} {\bibfnamefont {T.~P.}\ \bibnamefont {Simula}}, \ and\
  \bibinfo {author} {\bibfnamefont {P.~B.}\ \bibnamefont {Blakie}},\
  }\href@noop {} {\bibfield  {journal} {\bibinfo  {journal} {Phys. Rev. A}\
  }\textbf {\bibinfo {volume} {79}},\ \bibinfo {pages} {033626} (\bibinfo
  {year} {2009})}\BibitemShut {NoStop}%
\bibitem [{\citenamefont {Holzmann}\ \emph {et~al.}(2010)\citenamefont
  {Holzmann}, \citenamefont {Chevallier},\ and\ \citenamefont
  {Krauth}}]{holzmann}%
  \BibitemOpen
  \bibfield  {author} {\bibinfo {author} {\bibfnamefont {M.}~\bibnamefont
  {Holzmann}}, \bibinfo {author} {\bibfnamefont {M.}~\bibnamefont
  {Chevallier}}, \ and\ \bibinfo {author} {\bibfnamefont {W.}~\bibnamefont
  {Krauth}},\ }\href@noop {} {\bibfield  {journal} {\bibinfo  {journal} {Phys.
  Rev. A}\ }\textbf {\bibinfo {volume} {81}},\ \bibinfo {pages} {043622}
  (\bibinfo {year} {2010})}\BibitemShut {NoStop}%
\bibitem [{\citenamefont {Hadzibabic}\ \emph {et~al.}(2006)\citenamefont
  {Hadzibabic}, \citenamefont {Kruger}, \citenamefont {Cheneau}, \citenamefont
  {Battelier},\ and\ \citenamefont {Dalibard}}]{Hadzibabic}%
  \BibitemOpen
  \bibfield  {author} {\bibinfo {author} {\bibfnamefont {Z.}~\bibnamefont
  {Hadzibabic}}, \bibinfo {author} {\bibfnamefont {P.}~\bibnamefont {Kruger}},
  \bibinfo {author} {\bibfnamefont {M.}~\bibnamefont {Cheneau}}, \bibinfo
  {author} {\bibfnamefont {B.}~\bibnamefont {Battelier}}, \ and\ \bibinfo
  {author} {\bibfnamefont {J.}~\bibnamefont {Dalibard}},\ }\href@noop {}
  {\bibfield  {journal} {\bibinfo  {journal} {Nature}\ }\textbf {\bibinfo
  {volume} {441}},\ \bibinfo {pages} {1118} (\bibinfo {year}
  {2006})}\BibitemShut {NoStop}%
\bibitem [{\citenamefont {Chiao}\ \emph {et~al.}(2004)\citenamefont {Chiao},
  \citenamefont {Hansson}, \citenamefont {Leinaas},\ and\ \citenamefont
  {Viefers}}]{chiao2004}%
  \BibitemOpen
  \bibfield  {author} {\bibinfo {author} {\bibfnamefont {R.}~\bibnamefont
  {Chiao}}, \bibinfo {author} {\bibfnamefont {T.}~\bibnamefont {Hansson}},
  \bibinfo {author} {\bibfnamefont {J.}~\bibnamefont {Leinaas}}, \ and\
  \bibinfo {author} {\bibfnamefont {S.}~\bibnamefont {Viefers}},\ }\href@noop
  {} {\bibfield  {journal} {\bibinfo  {journal} {Phys. Rev. A}\ }\textbf
  {\bibinfo {volume} {69}},\ \bibinfo {pages} {063816} (\bibinfo {year}
  {2004})}\BibitemShut {NoStop}%
\bibitem [{\citenamefont {Bogoliubov}(1947)}]{bogoliubov1947}%
  \BibitemOpen
  \bibfield  {author} {\bibinfo {author} {\bibfnamefont {N.}~\bibnamefont
  {Bogoliubov}},\ }\href@noop {} {\bibfield  {journal} {\bibinfo  {journal} {J.
  Phys. (USSR)}\ }\textbf {\bibinfo {volume} {11}},\ \bibinfo {pages} {4}
  (\bibinfo {year} {1947})}\BibitemShut {NoStop}%
\bibitem [{\citenamefont {Berezinskii}(1971)}]{berezinskii}%
  \BibitemOpen
  \bibfield  {author} {\bibinfo {author} {\bibfnamefont {V.~L.}\ \bibnamefont
  {Berezinskii}},\ }\href@noop {} {\bibfield  {journal} {\bibinfo  {journal}
  {Soviet JETP}\ }\textbf {\bibinfo {volume} {32}},\ \bibinfo {pages} {493}
  (\bibinfo {year} {1971})}\BibitemShut {NoStop}%
\bibitem [{\citenamefont {Petrov}(2003)}]{petrov2003bose}%
  \BibitemOpen
  \bibfield  {author} {\bibinfo {author} {\bibfnamefont {D.}~\bibnamefont
  {Petrov}},\ }\emph {\bibinfo {title} {Bose-Einstein condensation in
  low-dimensional trapped gases}},\ \href@noop {} {Ph.D. thesis},\ \bibinfo
  {school} {Amsterdam} (\bibinfo {year} {2003})\BibitemShut {NoStop}%
\bibitem [{\citenamefont {Chaikin}\ and\ \citenamefont
  {Lubensky}(2000)}]{chaikin}%
  \BibitemOpen
  \bibfield  {author} {\bibinfo {author} {\bibfnamefont {P.~M.}\ \bibnamefont
  {Chaikin}}\ and\ \bibinfo {author} {\bibfnamefont {T.~C.}\ \bibnamefont
  {Lubensky}},\ }\href@noop {} {\emph {\bibinfo {title} {{Principles of
  condensed matter physics}}}}\ (\bibinfo  {publisher} {Cambridge University
  Press},\ \bibinfo {year} {2000})\BibitemShut {NoStop}%
\bibitem [{\citenamefont {Kosterlitz}\ and\ \citenamefont
  {Thouless}(1973)}]{kosterlitz}%
  \BibitemOpen
  \bibfield  {author} {\bibinfo {author} {\bibfnamefont {J.~M.}\ \bibnamefont
  {Kosterlitz}}\ and\ \bibinfo {author} {\bibfnamefont {D.~J.}\ \bibnamefont
  {Thouless}},\ }\href@noop {} {\bibfield  {journal} {\bibinfo  {journal} {J.
  Phys. C}\ }\textbf {\bibinfo {volume} {6}},\ \bibinfo {pages} {1181}
  (\bibinfo {year} {1973})}\BibitemShut {NoStop}%
\bibitem [{\citenamefont {Nelson}\ and\ \citenamefont
  {Kosterlitz}(1977)}]{nelson1977}%
  \BibitemOpen
  \bibfield  {author} {\bibinfo {author} {\bibfnamefont {D.~R.}\ \bibnamefont
  {Nelson}}\ and\ \bibinfo {author} {\bibfnamefont {J.~M.}\ \bibnamefont
  {Kosterlitz}},\ }\href@noop {} {\bibfield  {journal} {\bibinfo  {journal}
  {Phys. Rev. Lett.}\ }\textbf {\bibinfo {volume} {39}},\ \bibinfo {pages}
  {1201} (\bibinfo {year} {1977})}\BibitemShut {NoStop}%
\bibitem [{\citenamefont {Bishop}\ and\ \citenamefont
  {Reppy}(1978)}]{bishop1978}%
  \BibitemOpen
  \bibfield  {author} {\bibinfo {author} {\bibfnamefont {D.~J.}\ \bibnamefont
  {Bishop}}\ and\ \bibinfo {author} {\bibfnamefont {J.~D.}\ \bibnamefont
  {Reppy}},\ }\href@noop {} {\bibfield  {journal} {\bibinfo  {journal} {Phys.
  Rev. Lett.}\ }\textbf {\bibinfo {volume} {40}},\ \bibinfo {pages} {1727}
  (\bibinfo {year} {1978})}\BibitemShut {NoStop}%
\bibitem [{\citenamefont {Mondal}\ \emph {et~al.}(2011)\citenamefont {Mondal},
  \citenamefont {Kumar}, \citenamefont {Chand}, \citenamefont {Kamlapure},
  \citenamefont {Saraswat}, \citenamefont {Seibold}, \citenamefont {Benfatto},\
  and\ \citenamefont {Raychaudhuri}}]{mondal}%
  \BibitemOpen
  \bibfield  {author} {\bibinfo {author} {\bibfnamefont {M.}~\bibnamefont
  {Mondal}}, \bibinfo {author} {\bibfnamefont {S.}~\bibnamefont {Kumar}},
  \bibinfo {author} {\bibfnamefont {M.}~\bibnamefont {Chand}}, \bibinfo
  {author} {\bibfnamefont {A.}~\bibnamefont {Kamlapure}}, \bibinfo {author}
  {\bibfnamefont {G.}~\bibnamefont {Saraswat}}, \bibinfo {author}
  {\bibfnamefont {G.}~\bibnamefont {Seibold}}, \bibinfo {author} {\bibfnamefont
  {L.}~\bibnamefont {Benfatto}}, \ and\ \bibinfo {author} {\bibfnamefont
  {P.}~\bibnamefont {Raychaudhuri}},\ }\href@noop {} {\bibfield  {journal}
  {\bibinfo  {journal} {Phys. Rev. Lett.}\ }\textbf {\bibinfo {volume} {107}},\
  \bibinfo {pages} {217003} (\bibinfo {year} {2011})}\BibitemShut {NoStop}%
\bibitem [{\citenamefont {Prokof'ev}\ \emph {et~al.}(2001)\citenamefont
  {Prokof'ev}, \citenamefont {Ruebenacker},\ and\ \citenamefont
  {Svistunov}}]{prokofev2001}%
  \BibitemOpen
  \bibfield  {author} {\bibinfo {author} {\bibfnamefont {N.}~\bibnamefont
  {Prokof'ev}}, \bibinfo {author} {\bibfnamefont {O.}~\bibnamefont
  {Ruebenacker}}, \ and\ \bibinfo {author} {\bibfnamefont {B.}~\bibnamefont
  {Svistunov}},\ }\href@noop {} {\bibfield  {journal} {\bibinfo  {journal}
  {Phys. Rev. Lett.}\ }\textbf {\bibinfo {volume} {87}},\ \bibinfo {pages}
  {270402} (\bibinfo {year} {2001})}\BibitemShut {NoStop}%
\bibitem [{\citenamefont {Fisher}\ and\ \citenamefont
  {Hohenberg}(1988)}]{fisher1988}%
  \BibitemOpen
  \bibfield  {author} {\bibinfo {author} {\bibfnamefont {D.~S.}\ \bibnamefont
  {Fisher}}\ and\ \bibinfo {author} {\bibfnamefont {P.~C.}\ \bibnamefont
  {Hohenberg}},\ }\href@noop {} {\bibfield  {journal} {\bibinfo  {journal}
  {Phys. Rev. B}\ }\textbf {\bibinfo {volume} {37}},\ \bibinfo {pages} {4936}
  (\bibinfo {year} {1988})}\BibitemShut {NoStop}%
\bibitem [{\citenamefont {Bolda}\ \emph {et~al.}(2001)\citenamefont {Bolda},
  \citenamefont {Chiao},\ and\ \citenamefont {Zurek}}]{bolda}%
  \BibitemOpen
  \bibfield  {author} {\bibinfo {author} {\bibfnamefont {E.~L.}\ \bibnamefont
  {Bolda}}, \bibinfo {author} {\bibfnamefont {R.~Y.}\ \bibnamefont {Chiao}}, \
  and\ \bibinfo {author} {\bibfnamefont {W.~H.}\ \bibnamefont {Zurek}},\ }\href
  {\doibase 10.1103/PhysRevLett.86.416} {\bibfield  {journal} {\bibinfo
  {journal} {Phys. Rev. Lett.}\ }\textbf {\bibinfo {volume} {86}},\ \bibinfo
  {pages} {416} (\bibinfo {year} {2001})}\BibitemShut {NoStop}%
\bibitem [{\citenamefont {Leboeuf}\ and\ \citenamefont
  {Moulieras}(2010)}]{leb}%
  \BibitemOpen
  \bibfield  {author} {\bibinfo {author} {\bibfnamefont {P.}~\bibnamefont
  {Leboeuf}}\ and\ \bibinfo {author} {\bibfnamefont {S.}~\bibnamefont
  {Moulieras}},\ }\href {\doibase 10.1103/PhysRevLett.105.163904} {\bibfield
  {journal} {\bibinfo  {journal} {Phys. Rev. Lett.}\ }\textbf {\bibinfo
  {volume} {105}},\ \bibinfo {pages} {163904} (\bibinfo {year}
  {2010})}\BibitemShut {NoStop}%
\bibitem [{\citenamefont {Josephson}(1966)}]{Jos}%
  \BibitemOpen
  \bibfield  {author} {\bibinfo {author} {\bibfnamefont {B.~D.}\ \bibnamefont
  {Josephson}},\ }\href@noop {} {\bibfield  {journal} {\bibinfo  {journal}
  {Phys. Lett.}\ }\textbf {\bibinfo {volume} {21}},\ \bibinfo {pages} {608}
  (\bibinfo {year} {1966})}\BibitemShut {NoStop}%
\bibitem [{\citenamefont {Mandel}\ and\ \citenamefont {Wolf}(1995)}]{mandel}%
  \BibitemOpen
  \bibfield  {author} {\bibinfo {author} {\bibfnamefont {L.}~\bibnamefont
  {Mandel}}\ and\ \bibinfo {author} {\bibfnamefont {E.}~\bibnamefont {Wolf}},\
  }\href@noop {} {\emph {\bibinfo {title} {Optical coherence and quantum
  optics}}}\ (\bibinfo  {publisher} {Cambridge university press},\ \bibinfo
  {year} {1995})\BibitemShut {NoStop}%
\bibitem [{\citenamefont {Pitaevskii}\ and\ \citenamefont
  {Rosch}(1997)}]{Pitaevskii1997}%
  \BibitemOpen
  \bibfield  {author} {\bibinfo {author} {\bibfnamefont {L.~P.}\ \bibnamefont
  {Pitaevskii}}\ and\ \bibinfo {author} {\bibfnamefont {A.}~\bibnamefont
  {Rosch}},\ }\href@noop {} {\bibfield  {journal} {\bibinfo  {journal} {Phys.
  Rev. A}\ }\textbf {\bibinfo {volume} {55}},\ \bibinfo {pages} {R853}
  (\bibinfo {year} {1997})}\BibitemShut {NoStop}%
\bibitem [{\citenamefont {Bloch}\ \emph {et~al.}(2008)\citenamefont {Bloch},
  \citenamefont {Dalibard},\ and\ \citenamefont {Zwerger}}]{Bloch2008}%
  \BibitemOpen
  \bibfield  {author} {\bibinfo {author} {\bibfnamefont {I.}~\bibnamefont
  {Bloch}}, \bibinfo {author} {\bibfnamefont {J.}~\bibnamefont {Dalibard}}, \
  and\ \bibinfo {author} {\bibfnamefont {W.}~\bibnamefont {Zwerger}},\
  }\href@noop {} {\bibfield  {journal} {\bibinfo  {journal} {Rev. Mod. Phys.}\
  }\textbf {\bibinfo {volume} {80}},\ \bibinfo {pages} {885} (\bibinfo {year}
  {2008})}\BibitemShut {NoStop}%
\end{thebibliography}


%
\end{document}